\documentclass[preprint,showpacs,preprintnumbers,amsmath,amssymb,natbib]{revtex4}
\pdfoutput=1
\usepackage{graphicx}
\usepackage{epsfig}	
\usepackage{dcolumn}
\usepackage{bm}

\def\0{\mbox{\tiny $0$}}
\def\1{\mbox{\tiny $1$}}
\def\2{\mbox{\tiny $2$}}
\def\3{\mbox{\tiny $3$}}
\def\4{\mbox{\tiny $4$}}
\def\5{\mbox{\tiny $5$}}
\def\6{\mbox{\tiny $6$}}
\def\7{\mbox{\tiny $7$}}
\def\8{\mbox{\tiny $8$}}
\def\9{\mbox{\tiny $9$}}

\def\f14{\mbox{\tiny $\frac{1}{4}$}}

\newcommand{\st}{(x^i,p_j,\eta)}                                         
\newcommand{\ns}{(x^i,n_j,\eta)}                                         
\renewcommand{\d}[2]{\frac{\partial #1}{\partial #2}}                   
\newcommand{\pc}[1]{\left[#1\right]}                                     

\begin{document}

\title{C$\nu$B damping of primordial gravitational waves and the fine-tuning of the C$\gamma$B temperature anisotropy}
\author{A. E. Bernardini}
\email{alexeb@ufscar.br}
\author{J. F. G. Santos}
\email{jonas@df.ufscar.br}
\affiliation{Departamento de F\'{\i}sica, Universidade Federal de S\~ao Carlos, PO Box 676, 13565-905, S\~ao Carlos, SP, Brasil}

\date{\today}
\begin{abstract}

Damping of primordial gravitational waves due to the anisotropic stress contribution owing to the cosmological neutrino background (C$\nu$B) is investigated in the context of a radiation-to-matter dominated Universe.
Besides its inherent effects on the gravitational wave propagation, the inclusion of the C$\nu$B anisotropic stress into the  dynamical equations also affects the tensor mode contribution to the anisotropy of the cosmological microwave background (C$\gamma$B) temperature.
Given that the fluctuations of the C$\nu$B temperature in the (ultra)relativistic regime are driven by a multipole expansion, the mutual effects on the gravitational waves and on the C$\gamma$B are obtained through a unified prescription for a radiation-to-matter dominated scenario.
The results are confronted with some preliminary results for the radiation dominated scenario.
Both scenarios are supported by a simplified analytical framework, in terms of a scale independent dynamical variable, $k \eta$, that relates cosmological scales, $k$, and the conformal time, $\eta$.
The background relativistic (hot dark) matter essentially works as an effective dispersive medium for the gravitational waves such that the damping effect is intensified for the Universe evolving to the matter dominated era.
Changes on the temperature variance owing to the inclusion of neutrino collision terms into the dynamical equations result into spectral features that ratify that the multipole expansion coefficients $C_{l}^{T}$'s die out for $l \sim 100$.
\end{abstract}
\pacs{04.30.Db, 98.80.Hw}

\keywords{Gravitational Waves, Multipole Decomposition, Anisotropy Spectrum, Neutrinos}

\maketitle

\section{Introduction}

The theoretical investigation and the phenomenological analysis of anisotropies in the cosmological microwave background (C$\gamma$B) radiation is recursively considered as a singular valuable check on the validity of simple inflationary cosmological models.
The fast growing of primordial masses and of energy density fluctuations are identified as the simplest mechanism for producing cosmological structures and observable C$\gamma$B temperature anisotropies.
In addition, a primordial spectrum of gravitational waves \cite{Hanson,BICEP} may also have been perturbatively induced during the inflationary epoch.
It could, for instance, change the theoretical predictions for cluster abundances, and work as a pertinent test for inflationary models as it produces some imprints on radiation tensor modes.

Cosmological tensor fluctuations should produce not only temperature anisotropies but also distinct imprints in the so-called magnetic or $B$-modes of its polarization field \cite{Hu97}, which has been identified through the C$\gamma$B polarization experiments \cite{Hanson,BICEP,Tu05,Bennet}.
Current experiments have indeed been able to put upper limits on polarizations of the C$\gamma$B that might be owed to a gravitational wave background \cite{LIGO,VIRGO,GEO600,TAMA300,LISA}.

The observed pattern of temperature anisotropies, when combined with probes of inhomogeneities in matter on large scale structures, and with measurements of the total energy density in the Universe, is in striking agreement with the simplest predictions for the spectrum of anisotropies due to gravitational waves produced during inflation.
These facts support the inclusion of extra ingredients in the fine-tuning analysis involving the theoretical predictions and the observable data for C$\gamma$B anisotropies.

The C$\nu$B contribution to the dark matter inventory at present can be estimated from the modifications on the matter power spectrum, even for neutrinos behaving like a relativistic fluid at higher redshifts \cite{Bernardini2011}.
This phenomenological characteristic is related to the large scale structures, such that effective mass values for neutrinos through the C$\gamma$B results are inferred through the transfer function in the matter power spectrum at small scales \cite{Dodelson,Dolgov02,Bitten}.
Depending on the current thermodynamic regime, the free streaming massive neutrinos can affect the cosmological evolution of tensor modes by increasing the magnitude of the anisotropic stress, which acts as an effective viscosity, absorbing gravitational waves in the low frequency.
That is the theoretical point discussed in some previous issues \cite{Lattanzi1,Lattanzi2, Weinberg} where it has been given some emphasis on the cosmological evolution of perturbation tensor modes coupled to cosmological neutrinos in the radiation dominated (RD) universe.

Our aim is to extend such a preliminary approach involving the RD cosmic inventory to a transient, radiation-to-matter dominated (RMD) background universe.
We shall follow the analytical setup based on the multipole formalism that reproduces the procedure which deals with scalar perturbations \cite{Ma:1995ey,Bernardini2011,Lattanzi1,Lattanzi2}.
Even in the framework for a RMD scenario, it can be shown that equations can be manipulated in order to avoid explicit (and sometimes confused) dependencies on cosmological scales, $k$, which, in this case, are absorbed by the scale independent variable, $k \eta$.
Besides quantifying the dynamical evolution of gravitational waves and identifying the role of neutrinos and collision terms inherent to the model, one shall be able to quantify a modified tensor mode variance for the temperature anisotropy.
Once extended to the RMD background scenario, our analysis follows several theoretical prescriptions provided by some preliminary studies like in Refs.~\cite{Lattanzi1,Lattanzi2,Weinberg,Bernardini2011}.

Our manuscript is therefore organized as follows.
In section II, we report about the textbook multipole formalism \cite{Ma:1995ey}, with the corresponding modifications for reconstructing the pattern of tensor perturbations \cite{Lattanzi1,Lattanzi2}.
In section III, we reproduce the framework for including the anisotropic stress effects on the propagation of gravitational waves by assuming physically reliable conditions over the collision parameters.
The dynamical evolution of tensor modes and its corresponding potential modifications on the C$\gamma$B temperature for a RMD environment is therefore quantified.
Since the neutrino viscosity underlies an increasing wave damping effect, we expect to have a frequency-dependent absorption of gravitational waves in the frequency range where neutrino decoupling happens.
We draw our conclusions in section IV.

\section{Theoretical preliminaries}

As supported by the decomposition theorem \cite{Ma:1995ey,Pas06,Dodelson,Bernardini2011},
the perturbation equations for the cosmological scenario in the synchronous gauge allows one to depict simpler and clearer properties of cosmological tensor perturbations.
In general lines, the cosmological evolution of a homogeneous Friedmann-Robertson-Walker (FRW) flat universe with background energy density, $\bar{\rho}(\eta)$, and pressure, $\bar{\mathcal{P}}(\eta)$, is described in terms of the scale factor, $a(\eta)$, through the following components of the Einstein equation,
\begin{eqnarray}
\label{friedmann01}
\left( {da/d\eta \over a} \right)^{2} &=& {8\pi\over3}G a^{2} \bar{\rho} \,,\\
{d\over d\eta} \left( {da/d\eta \over a} \right)&=& -{4\pi\over 3}G a^{2} (\bar{\rho} + 3 \bar{\mathcal{P}}) \,,
\label{friedmann02}
\end{eqnarray}
where $\eta$ is the conformal time defined by $d\eta = dt / a$, $G$ is the Newtonian constant, and one sets the natural units $c = \hbar = k_B = 1$.

The propagation of gravitational waves is parameterized by the relevant (spatial) components of the perturbed metric written as \footnote{One notices the $(-+++)$ signature for the metric.}
\begin{equation}
g_{ij}= a^2(\eta)\left[\delta_{ij} + h_{ij}\right],\quad i,j =1,\,2,\,3,
\end{equation}
while $g_{00}=-1$ and $g_{0i}=0$.
The transverse traceless part of $h_{ij}$ corresponds to gravitational waves \cite{Dodelson}.
The coupling of tensor modes with matter and radiation is suppressed in case of perfect fluids.
However, the inclusion of traceless transverse terms, $\Pi^i_j$, into the anisotropic stress tensor, $T^{i}_{j}$, as defined by
\begin{equation}\label{T_ideal}
T^\mu_\nu=(\bar{\mathcal{P}}+\delta\bar{\mathcal{P}}) g^\mu_\nu+((\bar{\mathcal{\rho}}+\delta\bar{\mathcal{\rho}})+(\bar{\mathcal{P}}+\delta\bar{\mathcal{P}}))U^{\mu}U_{\nu},
\quad \mu,\nu =0,\,1,\,2,\,3,
\end{equation}
modifies the isotropic and homogeneous characteristics of perfect fluids, $\bar{\rho}$ and $\bar{\mathcal{P}}$, and provides a natural coupling for observing interactions between the tensor modes, i. e. gravitational waves, in the (RMD) cosmological environment.
It changes the dynamical behavior of the tensor perturbation components, $h_{ij}$, through the following equation of motion,
\begin{equation}
\partial^2_t{h}_{ij}+\left(\frac{3}{a}\frac{da}{dt}\right)\partial_t{h}_{ij}-\left
(\frac{\nabla^2}{a^2}\right)h_{ij}=16\pi G\frac{\Pi_{ij}}{a^2}.
\label{einst}
\end{equation}
In this case, the traceless component of the energy-momentum tensor is defined by
$\Pi_{ij} = g_{il}\Pi^l_j= g_{il}( T^l_j- \delta^l_j T_k^k/3) = T_{ij}- g_{ij} (\bar{\mathcal{P}}+\delta\bar{\mathcal{P}}) $.

By turning Eq. (\ref{einst}) into its Fourier space transformed form, one has
\begin{equation}
\ddot h_{ij} + 2{\mathcal H} \dot h_{ij} + k^2 h_{ij} =16\pi G \Pi_{ij},
\label{eq:einstK}
\end{equation}
where ${\mathcal H} = \dot a/a$, and {\em dots} correspond to conformal time derivatives.
The anisotropic stress $\Pi_{ij}$ is given by
\begin{equation}
\Pi_{ij}=T_{ij}-\frac{g_{ij}}{3}T^k_k =
\frac{a^2\bar\rho}{4\pi}\int  \left(n_i n_j -\frac{\delta_{ij}}{3}\right) F_\nu d\Omega
= \frac{a^2\bar\rho_\nu}{4\pi}\mathcal{F}_{ij}^{(0)},
\label{zeroth}
\end{equation}
with $\int n_i n_j d\Omega=4\pi \delta_{ij}/3$.

The analytical multipole decomposition discussed in Ref.~\cite{Lattanzi1} supports the equations for obtaining
$\mathcal{F}_{ij}^{(0)}$.
Given a (Fermi-Dirac) momentum distribution, $f_0(q)$, and scalar perturbations, $\Psi(k_i,\,q,\, n_j,\,\eta)$,
one defines
\begin{equation}
F_\nu(k_i,\, n_j,\,\eta) \equiv \frac{\int q^3 f_0(q) \Psi(k_i,\,q,\, n_j,\,\eta) dq}{\int q^3 f_0(q) dq},
\end{equation}
which appears into the Boltzmann equation \cite{Lattanzi1} as
\begin{equation}
\dot F_\nu + i\, k_i n^i F_\nu + 2\dot h_{ij}n^i n^j = \frac{4\pi}{a^4 \bar \rho_\nu}\int q^3 \hat C[f] dq,
\label{eq:boltzF}
\end{equation}
where the interactions brought up by $\hat C[f]$ shall be discussed latter.
By following the same notation from \cite{Lattanzi1}, one finds that
\begin{equation}
\mathcal{F}_{ij}(k_i,\,\mu,\,\eta) = \int_0^{2\pi} \left(n_i n_j-\frac{\delta_{ij}}{3}\right) F_\nu\, d\varphi,
\label{varphi}
\end{equation}
where $\varphi$ is the polar angle such that $d\Omega = \sin\theta d\theta\,d\varphi$.
Upon multiplying Eq.~(\ref{eq:boltzF}) by $(n_i n_j-\delta_{ij}/3)$ and integrating over $\varphi$, one has \cite{Bernardini2011,Lattanzi1},
\begin{equation}
\dot {\mathcal{F}}_{ij} + i k \mu\, \mathcal{F}_{ij} +2\dot h_{l m}\int_0^{2\pi} n^l n^m \left(n_i n_j-\frac{\delta_{ij}}{3}\right) d\varphi = {\mathcal C}_{ij}
\label{eq:boltz2}
\end{equation}
with $\mu \equiv \cos(\varphi) = \hat k \cdot \hat n$, where an ordinary collision term, ${\mathcal C}_{ij}$, is introduced as,
\begin{equation}
{\mathcal C}_{ij}\equiv\frac{4\pi}{a^4 \bar \rho}\int q^3 dq \int_0^{2\pi} d\varphi \, \left(n_i n_j -\frac{\delta_{ij}}{3}\right) \hat C[f].
\end{equation}
Now performing the Legendre expansion with respect to $\mu$, one obtains
\begin{align}
\mathcal{F}_{ij}(k_i,\mu ,\eta) &= \sum_{\ell=0}^\infty (-i)^{\ell}(2\ell+1)\mathcal{F}_{ij}^{(\ell)}(k_i,\eta)P_{\ell}(\mu),\\
\mathcal{C}_{ij}(k_i,\mu ,\eta) &= \sum_{\ell=0}^\infty (-i)^{\ell}(2\ell+1)\mathcal{C}_{ij}^{(\ell)}(k_i,\eta)P_{\ell}(\mu),
\end{align}
for which the orthogonality relations involving the Legendre polynomials,
\begin{equation}
\int_{-1}^{1} P_{\ell} P_{m}\,d\mu = \frac{2}{2\ell+1} \delta_{\ell m},
\end{equation}
are prescribed.
The zeroth-order multipole contribution that appears into Eq.~(\ref{zeroth}) is effectively the unique non-vanishing contribution of $F_{\nu}$ into Eq.~(\ref{varphi}), which is computed from the above multipole expansion for $\mathcal{F}_{ij}(k_i,\mu ,\eta)$.
It explains the origin of the anisotropic stress contribution written in terms of ${\mathcal{F}}_{ij}^{(0)}$ into Eq.~(\ref{einst}).

Finally, by multiplying Eq.~(\ref{eq:boltz2}) by $(i^\ell/2) P_{\ell}$ and integrating it over $\mu$, after performing some straightforward mathematical manipulations, one obtains \cite{Lattanzi1}
\begin{align}
\dot {\mathcal{F}}_{ij}^{(0)} &= - k\,\mathcal{F}_{ij}^{(1)} - \frac{8\pi }{15} \dot h_{ij } +\mathcal{C}_{ij}^{(0)}, \label{eq:bolt_mpol1} \\[0.5cm]
\dot {\mathcal{F}}_{ij}^{(2)} &= -\frac{k}{5} \left[3 \mathcal{F}_{ij}^{(3)} -  2 \mathcal{F}_{ij}^{(1)} \right] - \frac{16\pi}{105} \dot h_{ij }+\mathcal{C}_{ij}^{(2)},\\[0.5cm]
\dot {\mathcal{F}_{ij}}^{(4)} &= - \frac{k}{9} \left[   5 \mathcal{F}_{ij}^{(5)} -  4 \mathcal{F}_{ij}^{(3)} \right] - \frac{8\pi}{315} \dot h_{ij } +\mathcal{C}_{ij}^{(4)},\\[0.5cm]
\dot {\mathcal{F}_{ij}}^{(\ell)} &= -\frac{k}{2\ell+1} \left[   (\ell+1) \mathcal{F}_{ij}^{(\ell+1)} -  \ell\, \mathcal{F}_{ij}^{(\ell-1)} \right] +\mathcal{C}_{ij}^{(\ell)}\qquad (\ell \ne 0,2,4). \label{eq:bolt_mpol4}
\end{align}
The above equations are constrained by the dynamical behavior of $h_{ij}$, which turns them into a system of $\ell+1$ decomposed first-order ordinary differential equations completely equivalent to the Boltzmann equation.

Since we are concerned with the fact that the anisotropic stress is only cosmologically relevant for massless particles \cite{Pas06}, independently of our previous arguments, the condition of having background neutrinos in ultra-relativistic thermodynamic regime is assumed along the RMD era.
In this case one can write $\bar{\rho}_{\nu}$ in terms of the total energy density, $\bar{\rho}$, and of the rates 
$R_\nu =  \Omega_\nu/\Omega_r\equiv \bar{\rho}_{\nu}/\bar{\rho}_{r}$ and $R_{_{m/\gamma}} = \Omega_m/\Omega_\gamma$,
\begin{eqnarray}
\bar{\rho}_{\nu} &=& \bar{\rho}_{\nu}\frac{\bar{\rho}}{\bar{\rho}_m + \bar{\rho}_r} = 
\frac{1}{1+(\bar{\rho}_m/\bar{\rho}_r)}R_\nu \bar{\rho}=
\frac{1}{1+(\Omega_\gamma/\Omega_r)(\Omega_m/\Omega_\gamma)}R_\nu \bar{\rho}\nonumber\\
&=&\frac{1}{1+ (1-R_\nu)R_{_{m/\gamma}}}R_\nu \bar{\rho},
\label{ewse}
\end{eqnarray}
where $\bar{\rho}= \bar{\rho}_m + \bar{\rho}_r$, $\Omega_r = \Omega_{\nu} + \Omega_{\gamma}$, such that $\Omega_i = 3\rho_i/(8\pi G)$, with $i = r$ (radiation), $\nu$ (neutrinos), $\gamma$ (photons) and $m$ (matter).
By substituting Eq.~(\ref{zeroth}) with the above-defined parameters into Eq.~(\ref{einst}), and using Eq.~(\ref{friedmann02}) for $\bar{\rho}$, one obtains a suitably modified picture of Ref.~\cite{Lattanzi1} given by
\begin{equation}
\ddot h_{ij} + 2{\mathcal H} \dot h_{ij} + k^2 h_{ij} = \frac{3}{2\pi}\mathcal{H}^2\frac{R_\nu}{1+(1-R_\nu)R_{_{m/\gamma}}}\mathcal{F}_{ij}^{(0)},
\label{OHS}
\end{equation}
where the inclusion of the elements of the RMD cosmic inventory are evinced by $R_{_{m/\gamma}}$ on the right-hand side\footnote{By setting $R_{_{m/\gamma}} = 0$ one is able to recover the results for the RD cosmic inventory as in \cite{Lattanzi1}.}.

\section{Gravitational waves coupled to neutrinos in the RMD scenario}

The background solutions of the Friedmann equation for the RMD universe, with the corresponding equation of state respectively represented by $\mathcal{\bar {P}}_{r} = \bar\rho_{r}/3$ and $\mathcal{\bar P}_m = 0$ are given by
\begin{equation}
\label{rho(a)}
 \bar\rho_{r} = \rho_0 \frac{\Omega_{r}}{a^4}\,,
\qquad
 \bar\rho_m = \rho_0 \frac{\Omega_m}{a^3}\,,
\qquad
\end{equation}
where we have neglected the cosmological constant phase.
For a RMD cosmological background, the scale factor dependence on the conformal time reproducing the radiation-to-matter transition can be exactly given by
\begin{eqnarray}
a(\eta) &=& \frac{2\pi G}{3}\Omega_m\eta^2 + \left(\frac{8\pi G}{3}\Omega_{r}\right)^{1/2}\eta,
\end{eqnarray}
and conveniently re-written as
\begin{eqnarray}
k a(\eta) = \sqrt{\frac{8\pi G}{3}\Omega_{r}}\left[\frac{(k\eta)^2}{4 k\eta_{eq}} + k\eta\right],
\label{a(eta)}
\label{etate}
\end{eqnarray}
with $\eta_{eq} =(3\Omega_{r}/(8\pi G))^{1/2}/\Omega_{m}$, where a scale independent parameter $k\eta$ has been introduced, and the boundary conditions are set as $a(0) \equiv 0$.
Eq.~(\ref{etate}) also fiducially describes the dynamics deep inside radiation or matter dominated (MD) eras separately.
Assuming the above dependence of $a$ on $\eta$ into the coupled equations of the previous section, one can treat the gravitational waves entering the horizon even after the time of matter-radiation equality, i. e. with the redshift $z < 10^4$.
Therefore, besides being applied to the analysis of waves that have entered the horizon well inside the RD era, with $k \gg 0.1 [Mpc]^{-1}$, our result can be extended to the analysis of waves with $k \sim 0.1 [Mpc]^{-1}$.
In addition, from the point of view of the mathematical manipulation/resolution of the equations, the explicit dependence on the cosmological scales, $k$, will be relegated to the parameter $k \eta_{eq}$ at Eq.~(\ref{a(eta)}), so that one can express  all the subsequent results in terms of $k\eta$, with $\eta$ in units of $\eta_{0} \approx 1/H_{0} \approx 5000 [Mpc]$.
In this case, $k\eta \sim 1$ corresponds to the horizon crossing parameter.

The system of coupled equations ordinarily defined in terms of $k\eta$ also allows one to depict the behavior of waves deep inside the horizon $(k\eta \gg 1)$.
Given the scale covariance introduced by $k\eta$ (in place of a factorized dependence on $\eta$), it is always possible to re-scale the initial value of $h_{ij}(k\eta)$ as to have $h_{ij}(0) = h^{(0)}$.
The choice of the initial amplitude after crossing the causal horizon, $h^{(0)}$, is arbitrary and it does not affect our results.
The variables $\dot{h}_{ij}$ and $\mathcal{F}_{ij}^{(\ell)}$ are concerned with the information about the effect of damping oscillation.
It is attributed to the expansion of the Universe, in case of $\dot{h}_{ij}$, and to the interaction with the $C\nu B$, in case of  $\mathcal{F}_{ij}^{(\ell)}$. 
Therefore, these terms should be taken into account just after crossing the causal horizon, i. e. for $k\eta > k\eta_{0} \approx 0$.
Because of that, it is reasonable to assume $\dot{h}_{ij}(0) = 0$ and $\mathcal{F}_{ij}^{(\ell)}(0) = 0$. 
Finally, we also have assumed the standard cosmological values for $\Omega_i$, with $i = \gamma,\, \nu$ and $m$, such that $\Omega_\nu/\Omega_\gamma > 0$ and $\Omega_\gamma/\Omega_m \approx 10^{-4}$.

The corresponding dynamical evolution of the gravitational waves, i. e. of the tensor modes, $h_{ij}$, in terms of the scale independent variable, $k \eta$, can be depicted in Figs.~\ref{gw006a1} and \ref{gw006a2}.
Once gravitational waves have entered the horizon ($k\,\eta \gtrsim 1$), their amplitude dies away (c. f. Fig.~\ref{gw006a1}) more rapidly at a universe with the cosmic inventory containing the matter component contribution.
By suppressing the contribution due to the neutrino anisotropic stress at Eq.~(\ref{OHS}), one recovers a damped harmonic oscillator (DHO)-like equation for which the damping factor is give by $\gamma \equiv 2{\mathcal H}$.
One can notice that deep inside the MD era the $\gamma$ factor is two times the value corresponding to that of deep inside the RD era.
The amplitude of the gravitational waves are relatively suppressed when it penetrates into the MD era.

The neutrino free-streaming regime is obtained by setting a vanishing collision term, $\mathcal{C}_{ij}^{(\ell)}=0$, at the evolution equations.
Fig.~\ref{gw006a1} shows the results for $R_{\nu} = 0.4052$ (three neutrino species) and $R_{\nu} \approx 1$ (not so realistic large number of neutrino degrees of freedom, which includes extra flavor quantum numbers) obtained from numerical calculations involving $1200$ multipoles for RD and RMD scenarios.
The results are relevant for modes which enter the horizon at the Universe's temperature about $T \lesssim 1\,MeV$ \cite{Weinberg}.
The anisotropic stress effects are relatively suppressed for modes which enter the horizon at the MD era, as one can observe from the right side of Eq.~(\ref{OHS}).
The exception is for the situation where $R_{\nu}$ approximates to unity.
The amplitude $h_{ij}$ is constant outside the horizon and starts decreasing after the horizon crossing.

The inclusion of the matter background into the cosmic inventory introduces an additional subtle effect on the amplitude of the gravitational waves under the influence of the anisotropic stress.
Scales just entering the horizon at late times have the corresponding oscillation modes undergoing a delayed suppression due to the coupling to neutrinos.
It propagates to the following oscillation peaks in a kind of translational effect of the oscillation pattern, which is naturally expected if one observes that, in the limit of radiation domination, one has
\begin{equation}
|h_{ij}(k\eta)| = \frac{\sin{(k\eta)}}{k\eta},
\label{RDRD}
\end{equation}
and, in the limit of matter domination, one has
\begin{equation}
|h_{ij}(k\eta)| = 3\frac{\left(\sin{(k\eta)} - k\eta \,\cos{(k\eta)}\right)}{(k\eta)^3}.
\label{RDRDM}
\end{equation}
The reason for such a behavior is engendered by the fact the RMD curves are correctly interpreted only for $k \eta_{eq} = 1$, which correspond to scales that have entered the Hubble horizon at the time of matter-radiation equality.
In the Appendix we show the corresponding results for $k \eta_{eq} = 100$.
In spite of being evinced for the RMD scenario, increasing values of $R_{\nu}$ result into a more relevant suppression of the tensor modes even for the RD scenario.

\begin{figure}
\centering
\includegraphics[scale=1]{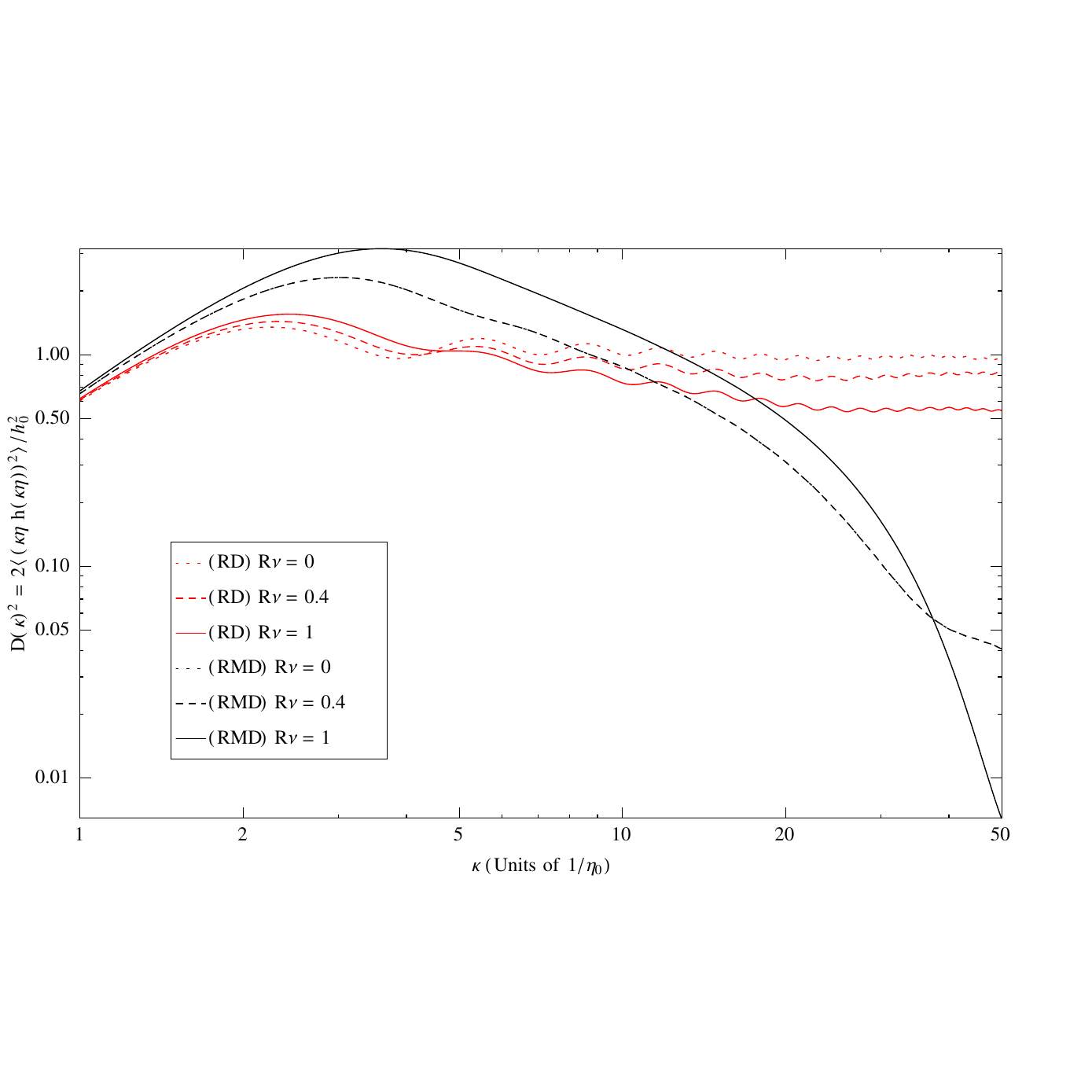}
\caption{Time-averaged values of $2 \langle(k\eta)^2|h_{ij}(k\eta)|^2\rangle$ as function of $k\,[Mpc^{-1}]$ for RD (red lines) and RMD (black lines) background cosmic inventories.
Results are for vanishing anisotropic stress, with $R_{\nu}=0$ (dotted lines), for $R_{\nu}=0.4 (0.4052)$ (dashed lines), and for $R_{\nu} = 1$ (solid lines).
Notice the {\em fake} resonance effect for modes with $k\eta\gtrsim 1$ followed by the neutrinos damping effect which is more relevant for scales deep inside the horizon.}
\label{gw006b1}
\end{figure}

In Ref.~\cite{Lattanzi1} one identifies that for the standard case corresponding to $R_{\nu} = 0.4052$ for three families of neutrinos, roughly $22\%$ of the intensity of the gravitational waves is absorbed by the $C\gamma B$ environment.
In spite of not considering the same steps for numerical integrations as assumed in Ref.~\cite{Lattanzi1}, 
our results for the RD era agrees with those presented in Ref.~\cite{Lattanzi1}, as it can be depicted in Figs.~\ref{gw006a1} and \ref{gw006b1} by comparing dashed and dotted red lines.
The effective suppression due to the inclusion of neutrinos can be depicted from Fig.~\ref{gw006b1} where we have computed the time-averaged quantity  $D(k)^2 = \langle 2 (k\eta)^2|h_{ij}(k\eta)|^2\rangle$ as function of the cosmological scale, $k$.
Such a time-averaged quantity is processed from a cut-off $\eta^*$.
Although $\eta^*$ is arbitrary, the time-averaged operation over $(k\eta)^2|h_{ij}(k\eta)|^2$ is effective only for scales entering the horizon at times $\eta \gg \eta^*$ .

When the elements for describing the RMD regime are introduced, our results are considerably different, in spite of exhibiting a conceptual agreement with those from Ref.~\cite{Lattanzi1}.
The point is that: once the cosmic inventory enters into the MD era, the effects due to anisotropic stress over the corresponding gravitational wave modes are highly suppressed.
There is an expected overall suppression of the gravitational wave modes (c. f. the black lines depicted in Figs.~\ref{gw006a1} and \ref{gw006b1}) driven by the MD regime.
Likewise, given that the realistic neutrino effects are suppressed, in the RMD scenario the the relative rate of absorption of waves turns into a tiny value $\ll 0.1\%$ (c. f. the dashed and dotted overlapping black lines depicted in Figs.~\ref{gw006a1} and \ref{gw006b1}).
One can notice that the lines obtained for $R_{\nu} \approx 1$ overpass the lines obtained for $R_{\nu} \approx 0$ at some ordinary scale $\tilde{k}$.
Scale values for which $k > \tilde{k}$ have tried out a sufficient number of oscillating cycles to average $h_{ij}(k\eta)$ and produce some representative damping effect.
In this case, the realistic effects produced by neutrinos correspond to a suppression of the power spectrum of gravitational waves for which $k \gg \tilde{k}$.

The maximum amount of damping occurs for the extrapolating limit of $R_\nu \rightarrow 1$.
In this case, the influence of matter on the cosmic inventory ($R_{m/\gamma} \sim 10^4$) is highly suppressed from Eq.~(\ref{ewse}) and therefore the damping effect increases (c. f. solid black lines depicted in Figs.~\ref{gw006a1} and \ref{gw006b1}).
Such a pictorial situation results into an unrealistic scenario for which, however, the absorption rate is roughly similar to that of $43\%$ from Ref.~\cite{Lattanzi1} (c. f. black lines crossing red lines in Fig.~\ref{gw006b1}).

Finally, the highest first oscillation peak for the RMD results depicted in Fig.~\ref{gw006b1} appears because of the abovementioned relative delay (phase difference) of the first oscillation damping of the gravitational waves in the RMD era, as depicted in  Fig.~\ref{gw006a1}, and supported by Eqs.~(\ref{RDRD}-\ref{RDRDM}).
Even creating a kind of horizon crossing {\em fake}-resonance effect, it disappears along the cosmological $\eta$ evolution.
Moreover, the increasing damping caused by the anisotropic stress of the standard (three family) neutrinos is much more effective at the RD regime.

Turning back to the collision term contributions, we shall follow the parametrization from Ref. \cite{Lattanzi1} that sets $\hat C[f] = - f_0\Psi /\tau$, where $\tau$ is the mean time between collisions.
In this case one has $\mathcal{C}_{ij}^{(\ell)}=-\mathcal{F}_{ij}^{(\ell)}/\tau$.
The auxiliary parameter in defining the strength of the interactions, $k \tau$, corresponds to the ratio between the wave frequency and the collision frequency.
One can compare the effects of including the collision term parameterized by $\tau = 0.01,\, 0.1,\, 1$ and $10$ in Figs.~\ref{gw006a2} and \ref{gw006b2}.
One should notice that the inclusion of collision effects parameterized by $k \tau$ at $\mathcal{C}_{ij}^{(0)}$ into Eq.~(\ref{eq:bolt_mpol4}) affects the gravitational wave evolution in a very subtle way.
Since one has $\tau$ in units of $\eta_{0}$ and $k$ in units of $1/\eta_{0}$, upon setting $\tau > 10 \gg 1$ one recovers the free-streaming (collisionless) results.
Otherwise, small values for $k \tau$ would correspond to very frequent collisions that dominate the dynamical evolution described by Eqs.~(\ref{eq:bolt_mpol1}-\ref{eq:bolt_mpol4}).
It results into $\mathcal{F}_{ij}\propto e^{-\eta/\tau}$, which leads to an exponential decay suppression of the anisotropic stress.
Decreasing values of $\tau$ therefore represents increasing collision rates, and consequently a less dispersive environment/effect due to the anisotropic stress.
Although strong deviations from the standard scenario with $R_\nu=0.4052$ are unlikely, damping effects as those obtained for $R_\nu=1$ become effective just when neutrinos enter the free-streaming regime and can be interpreted either as the existence of additional neutrino degrees of freedom (d.o.f) or as the existence of exotic fluid/particles in the early Universe.

Figs.~\ref{gw006a2} and \ref{gw006b2} also show that, for $k \gg \tilde{k}$, the amount of damping with respect to the vanishing stress contribution in case of RMD era is not regular.
It corresponds to a scale dependent effect.
The effects of increasing the frequency of the collisions by diminishing $\tau$ can be, at least superficially, quantified.
Is important to notice that either in the limit of radiation domination (c. f. Eq.~(\ref{RDRD})), or in the limit of matter domination (c. f. Eq.~(\ref{RDRDM})), where
\begin{equation}
|h_{ij}(k\eta)| \approx \frac{\sin{(k\eta)}}{k\eta}\left(1 + \mathcal{O}(k\eta)^2\right),
\end{equation}
scales just entering the horizon leads to non-decaying values for $D(k)^2$.
By following the same analogy with a DHO, it is also relevant to notice that $D(k)^2$ parameterizes the damping of the averaged valued of the DHO energy.
In fact,
\begin{equation}
(k\eta)^2\,|h_{ij}(k\eta)|^2 \propto k^3 \,P_h(k),
\end{equation}
where $P_h(k)$ is the power spectrum related to tensor modes \cite{Dodelson}.

Fig.~\ref{gw006b2} shows the time-averaged quantity, $D(k)^2$, by considering the effective collisions parameterized by $k \tau = 0.01$ and $10$. The effect of rare collisions is recovered for $k\tau \sim 10$.
As in Fig.~\ref{gw006b1}, it is possible identify the crossing value of $\tilde{k}$ for which the correct interpretation of $D(k)^2$ is pertinent.

To end up, the damping of gravitational waves also affects some spectral features related to the tensor contribution to the anisotropy spectrum.
From the analytical multipole decomposition \cite{Dodelson}, the contribution to the $C_l^T$'s can be written as
\small
\begin{equation}
C^T_{l,i} = \frac{(l-1)l(l+1)(l+2)}{\pi}\int^\infty _0{ dk \, k^2 \left|\frac{\Theta^T _{l-2,i}}{(2l-1)(2l+1)}+2\frac{\Theta^T _{l,i}}{(2l-1)(2l+3)}+\frac{\Theta^T _{l+2,i}}{(2l+1)(2l+3)}\right|^2},
\label{Eq8.93}
\end{equation}
\normalsize
where $i$ denotes $+$ and $\times$ modes.
$\Theta_{l,i}^T$ is obtained through
\begin{eqnarray}
\Theta^T _{l,i} &=& -\frac{1}{2}\int^{\eta _0} _{\eta^*} d\eta \,j_l[k(\eta_0 - \eta)]\dot{h}_{ij}(k,\eta),\nonumber\\
\Theta^{T^{\mbox{\tiny(MD)}}}_{l,i}&\approx& -\frac{1}{2}\int^{\eta _0} _{\eta^*} d\eta \,j_l[k(\eta_0 - \eta)]\frac{d}{d\eta}\left[\frac{3j_1(k\eta)}{k\eta}\right](P_h(k))^{1/2},\end{eqnarray}
where the last step stands for the analytical approximation for the MD scenario.
In this case, the departing amplitude of the gravitational waves is given in terms of $P_h^{1/2}$.

Substituting the results for $\Theta^T _{l}$ into Eq.~(\ref{Eq8.93}) allows one to compute the tensor imprints on the map of the C$\gamma$B temperature.
After some mathematical manipulations \cite{Dodelson}, the analytical expression for the RD scenario results into
\begin{align}
C^T_l &= 2\frac{9(l-1)l(l+1)(l+2)}{4\pi}\int^{\infty}_0 dk\, k^2 P_h(k)\nonumber\\
&\times\left| \int^{\eta_0}_0 d(k\eta)\frac{j_2(k\eta)}{k\eta}\left[\frac{j_{j-2}(k[\eta_0 - \eta])}{(2l-1)(2l+1)}+2\frac{j_l(k[\eta_0 - \eta])}{(2l-1)(2l+3)}+\frac{j_{l+2}(k[\eta_0 - \eta])}{(2l+1)(2l+3)}\right]\right|^2 ,
\label{Eq.A106}
\end{align}
where we have set the lower limit on the time integral equals to zero since the time $\eta^*$ at which the modes enter the horizon is assumed to satisfy $\eta^* \ll \eta_0$.
We have identically followed the approximations set by \cite{Dodelson}.
Since one has
\begin{equation}
P_h(k) = \frac{8\pi}{k^3}\frac{H^2}{m_{Pl}^2},
\end{equation}
by defining novel integration variables $y \equiv k\eta_0$ and $x\equiv k\eta$, one gets the analytical form given by
\begin{align}
C^T_l &= 36\left(\frac{H_{inf}}{m_{Pl}}\right)^2(l-1)l(l+1)(l+2)\int^\infty _0 \frac{dy}{y}\nonumber\\
&\times\left| \int^y _0 dx \frac{j_2(x)}{x} \left[\frac{j_{l-2}(y-x)}{(2l-1)(2l+1)}+2\frac{j_l(y-x)}{(2l-1)(2l+3)}+\frac{j_{l+2}(y-x)}{(2l+1)(2l+3)}\right]\right|^2,
\label{Eq.A107}
\end{align}
where $H_{inf}$ is the Hubble rate when the modes crossed the horizon (when $k\eta = 1$ early on), after being modulated by some transfer function that connects MD to RD scenarios \cite{Dodelson}.

The numerical results obtained for the tensor modes, $h_{ij}(k,\eta)$, allows one to compute the neutrino and collision effect imprints on the map of C$\gamma$B temperature in the RMD scenario.

\begin{figure}
\centering
\includegraphics[scale=1]{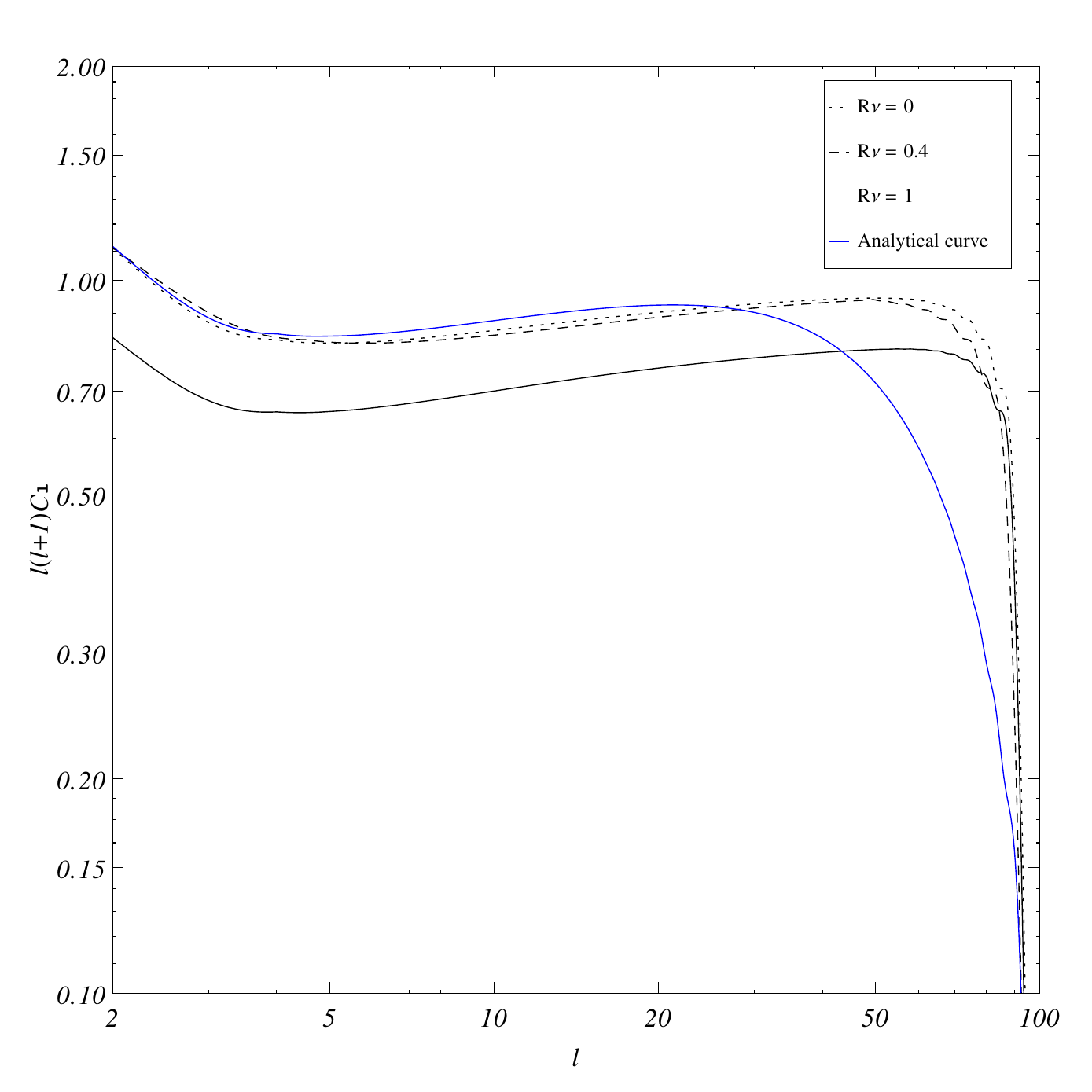}
\caption{Tensor contribution to the angular power spectrum in the RMD scenario.
Results are for vanishing anisotropic stress (dotted lines), for $R_{\nu}=0.4 (0.4052)$ (dashed lines), and for $R_{\nu} = 1$ (solid lines) for the collisionless case.
The blue line corresponds to the analytical results obtained for a vanishing anisotropic stress component, where we have  used an analytical transfer function to account for RD and MD scenarios.}
\label{gw007a}
\end{figure}

\begin{figure}
\centering
\includegraphics[scale=1]{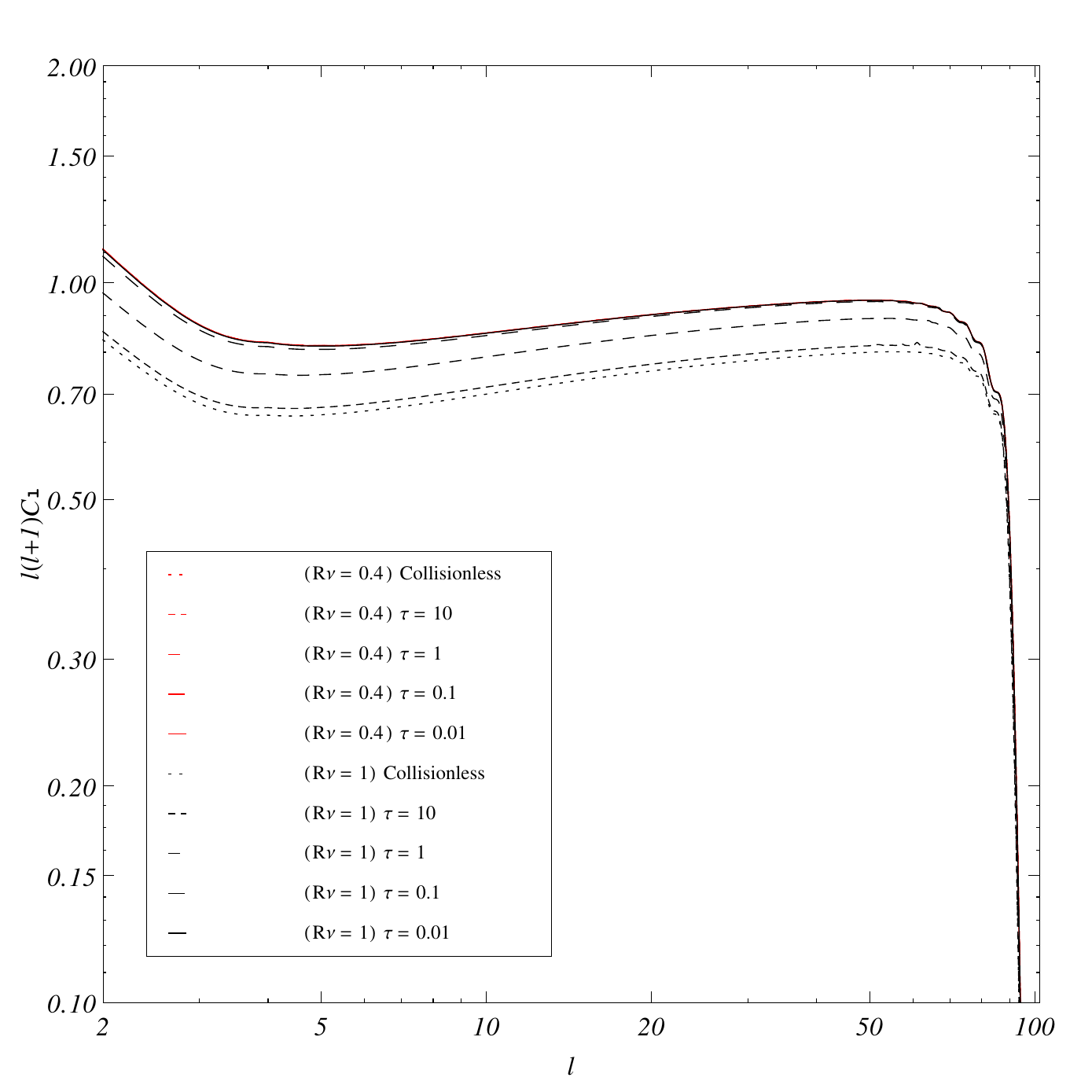}
\caption{Tensor contribution to the angular power spectrum for the RMD scenario in case of including the collision terms. Results are for $\tau = 0.01,\, 0.1,\, 1,$ and $10$, in units of $\eta_{0}$, and for the collisionless case.
Notice that, for decreasing values of the $\tau$ parameter, the suppression due to a huge number of neutrinos (d.o.f.) is not so relevant, i. e. the increasing collision effect attenuates the damping effect on the tensor mode propagation and it is reflected on the tensor contribution to the angular power spectrum.}
\label{gw007b}
\end{figure}

After entering the horizon, the amplitude of gravitational waves dies away (c. f. Fig.~\ref{gw006a1}).
The anisotropy spectrum are consequently affected by gravitational waves only on scales larger than the horizon at recombination.
This corresponds to angular scales $l \lesssim 100$ in the multipole expansion.
The tensor curves that we have obtained in Fig.~\ref{gw007a} for the same set of parameters introduced into Fig.~\ref{gw006a1} in the RMD era, show that $C_l^T$'s die out after $l \gtrsim 100$.
The analytical curve is obtained for RD connected to MD scenarios through a transfer function \cite{Turner}.
The coupling to neutrinos suppresses the contribution of tensor modes from the sum of anisotropies.
Therefore, if tensor perturbations grow up during the inflationary era, and if the total scalar plus tensor anisotropy spectrum is fit to the large-scale structure data, then the small-scale scalar amplitude is smaller than it would be.
The presence of the coupled anisotropic stress of neutrinos just shows the effects of the neutrinos (like a fluid) on the variance of temperature due to gravitational waves.
In Fig.~\ref{gw007b} we compare the effects of including the collision term parameterized by $k \tau = 0.01,\, 0.1,\, 1$ and $10$ assuming three neutrino species, $R_{\nu} = 0.4052$, and a larger number of neutrino species, $R_{\nu} = 1$.
From Fig.~\ref{gw006a2} one just notices some smooth suppression relative to decreasing of the frequency of collisions ($\tau \gg 1$).
The collision frequency also diminishes as like neutrinos go deep inside the free-streaming propagation regime, which intensifies the damping effect.

Fig.~\ref{gw007a} shows that for neutrinos with only three flavor degrees of freedom, the modifications on the $C^T_l$ coefficients are minimal.
The damping of the angular power spectrum and, more properly, of the temperature variance, $C^T_2$, are more relevant for $R_{\nu} \lesssim 1$.
The analytical curve reproduces the numerical results up to $l\sim30$.
For larger multipole values, with $l\gg100$, even numerically, the coefficients $C^T_l$ die out.
The sharp fall observed in Figs.~\ref{gw007a} and ~\ref{gw007b} is consistent with the multipole decomposition solution.

\section{Conclusions}

The observation of primordial gravitational waves indeed provides a renewed overview about the earliest moments in the history of the Universe and on possible new physics at energies many orders of magnitude beyond those accessible at particle accelerators.
The recent positive fit-back from experimental physics \cite{Hanson,BICEP} has indeed provided a crucial evidence for inflation in the early universe, which can also constrain the physics from the grand unification scale to the Planck scale.

Since a Universe overfilling viscosity results into gravitational wave damping effects, we have considered the possibility of observing some frequency-dependent absorption in the frequency range where neutrino decoupling is relevant.
By mixing analytical and numerical procedures, we have obtained the evolution of tensor modes and its corresponding imprints on the C$\gamma$B temperature in case of considering a RMD environment in the presence of an overfilling $C\nu B$.
Departing from the evolution of the gravitational waves from the time of their production, transversing the RD, the relevant modes exhibit a substantial damping on their amplitudes attributed to the expansion of the Universe when they enter into the MD era.
Meanwhile, the anisotropic stress component of the energy-momentum tensor changes the wave pattern when the cosmological neutrino background C$\nu$B is taken into account \cite{Xi08}.
It has been noticed that the effective neutrino viscosity introduces some increasing contribution to the overwhelming dynamics during the decoupling period.

The damping effects owing to the influence of the (neutrino) anisotropic stress have been computed for a RMD scenario and compared to previous results for the RD scenario.
We have compared the effects of including collision terms with collision frequency parameterized by $\tau = 0.01,\,0.1,\, 1,\,$ and $10$ assuming three neutrino species ($R_{\nu} = 0.4052$), and a larger number of (neutrinos) degrees of freedom ($R_{\nu} = 1$).
The collision dynamics is shown to introduce a tiny shift between two successive peaks of the gravitational wave spectrum.

The connection between the anisotropic stress of neutrinos and its effects on the C$\gamma$B temperature has also been identified, as it is used to be intermediated by gravitational waves.
Our results suggest that an extra number of neutrino degrees of freedom might be related either to some exotic neutrino family or even to some arbitrary composing contribution to the anisotropic stress.
In fact, considering $R_\nu = 1$ has intensified the damping effects up to its maximal value, as depicted in the map of the tensor contribution to the angular power spectrum, $C_l^T$.
For decreasing values of the collision $\tau$ parameter, the suppression of the gravitational wave amplitudes due to a huge number of degrees of freedom related to neutrinos is not so relevant, and the increasing collision frequency attenuates the damping effect on the tensor mode propagation.
It has also been reflected on the map of tensor contributions to the angular power spectrum.

Finally, a time-averaged quantity, $D(k)^2$, introduced to implicitly quantify the power spectrum of gravitational waves has shown that the RMD environment reduces the damping effect for realistic three flavor neutrino scenarios in spite of exhibiting the same maximal rate of damping for hypothesized scenarios with $R_{\nu} = 1$: a relevant aspect which may be considered in improving the computer programs used to analyze the future facilities.

\section*{Appendix}

\begin{figure}[center]
\centering
\includegraphics[scale=1]{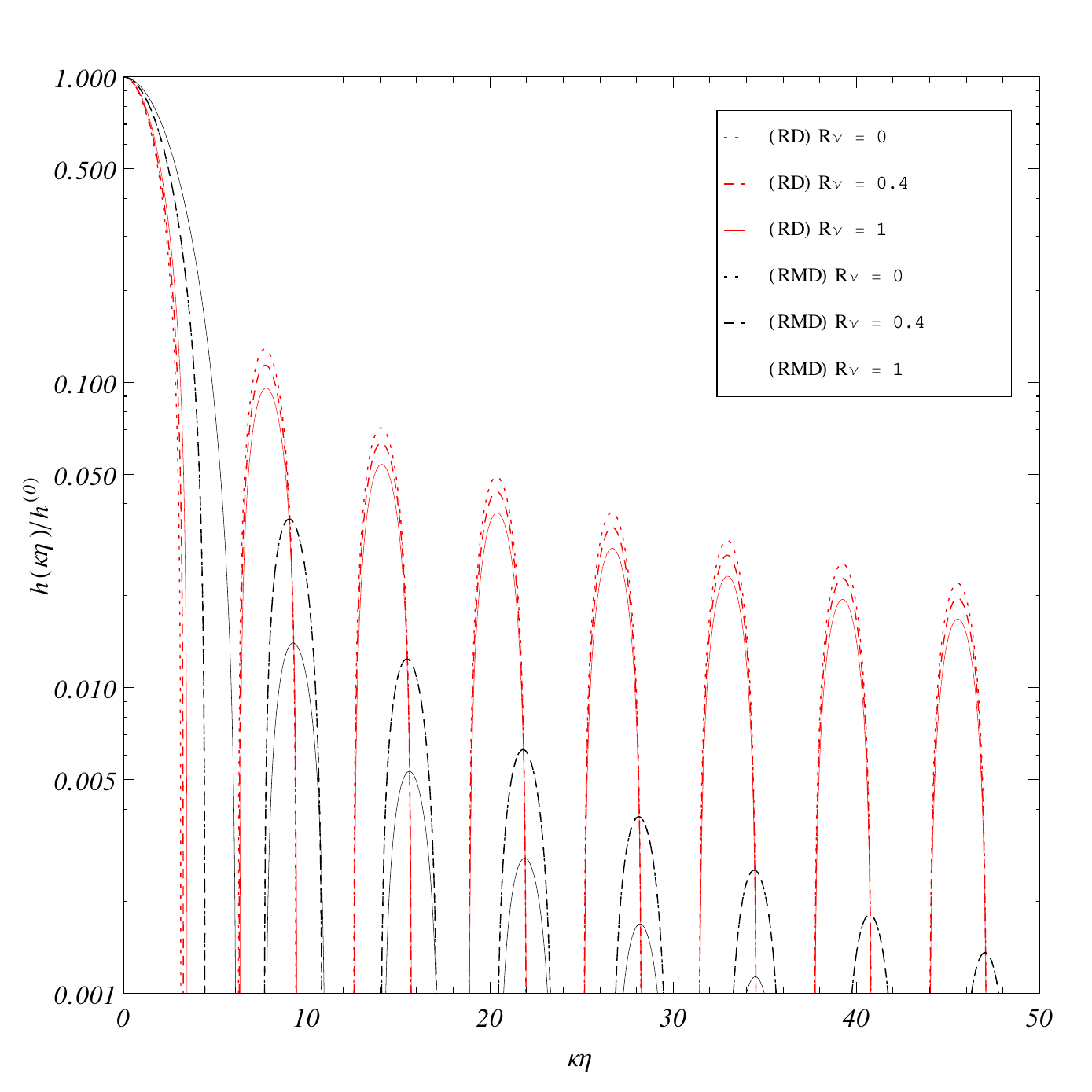}
\caption{Evolution of the normalized wave amplitude $h_{ij}/h_{ij}^{(0)}$ as function of $k \eta$ for RD (red lines) and RMD (black lines) background cosmic inventories.
Results are for vanishing anisotropic stress, with $R_{\nu}=0$ (dotted lines), for $R_{\nu}=0.4052$ (dashed lines), and for $R_{\nu} = 1$ (solid lines).
The RD curves are scale independent so that $k \eta$ is given in units of $k \eta_{0}$.
The RMD curves are correctly interpreted by observing that $k \eta_{eq} = 1$, which correspond to scales that have entered the Hubble horizon at the time of matter-radiation equality (See also Fig.~\ref{gw006a1C} for comparison).
In spite of being more evident for the RMD scenario, in both situations the largely increasing values of $R_{\nu}$ results into a more relevant suppression of the tensor modes during the cosmological evolution.
Notice that just for the first peak, solid lines are overpassing dashed- and dotted-lines.}
\label{gw006a1}
\end{figure}

\begin{figure}
\centering
\includegraphics[scale=1]{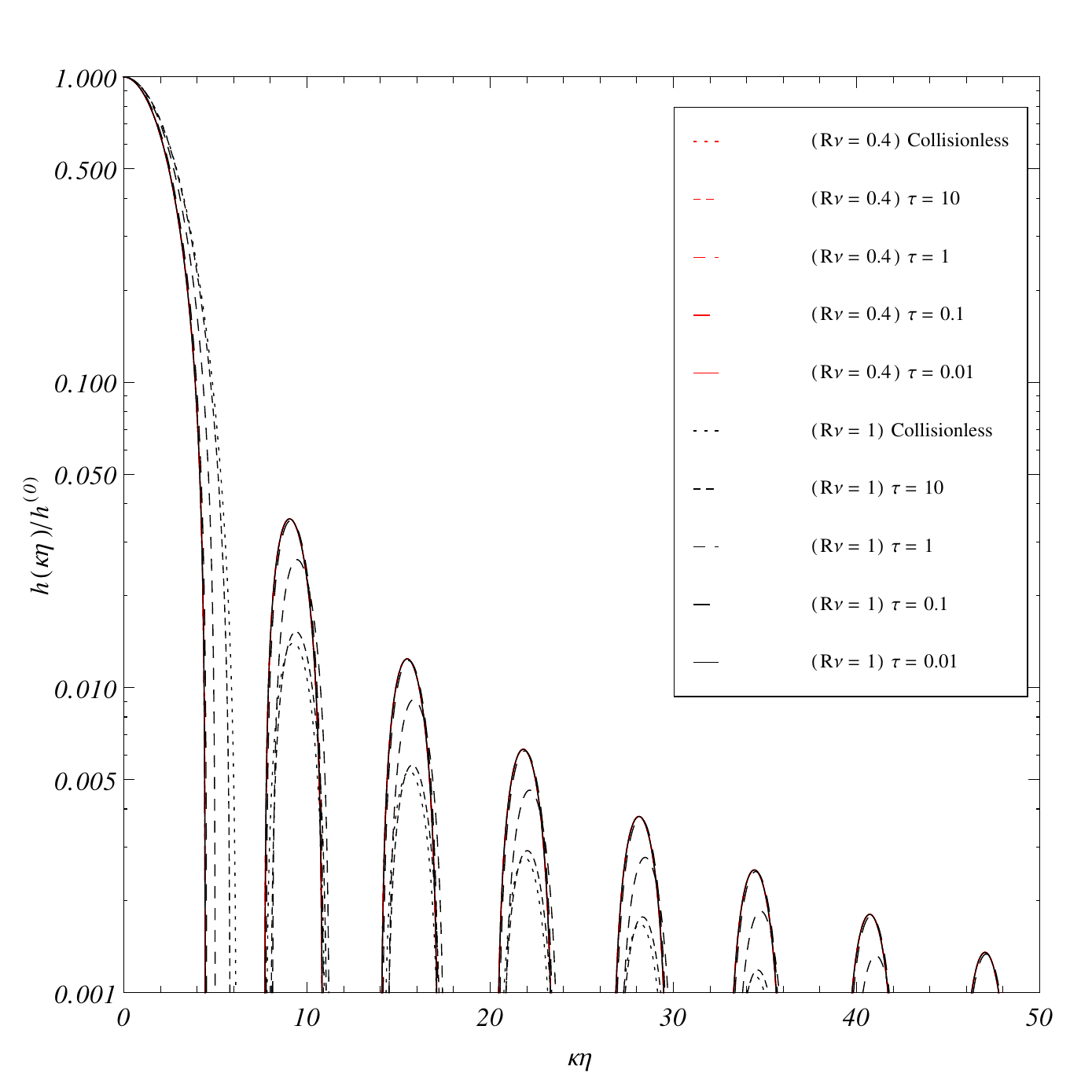}
\caption{Evolution of the normalized wave amplitude $h_{ij}/h_{ij}^{(0)}$ as function of $k \eta$ for the RMD scenario in case of including the collision terms.
Results are for $\tau = 0.01,\, 0.1,\, 1,$ and $10$, and for the collisionless case.
$\tau$ is given in units of $\eta_{0}$ and $k$ in units of $1/\eta_{0}$.
We have considered three neutrino species with $R_{\nu} = 0.4052 (0.4)$ (red lines) and the extreme case of a huge number of neutrino species with $R_{\nu} = 1$ (black lines).
The RMD curves are correctly interpreted by observing that $k \eta_{eq} = 1$ (see also Fig.~\ref{gw006a2C} for comparison).}
\label{gw006a2}
\end{figure}

\begin{figure}
\centering
\includegraphics[scale=1]{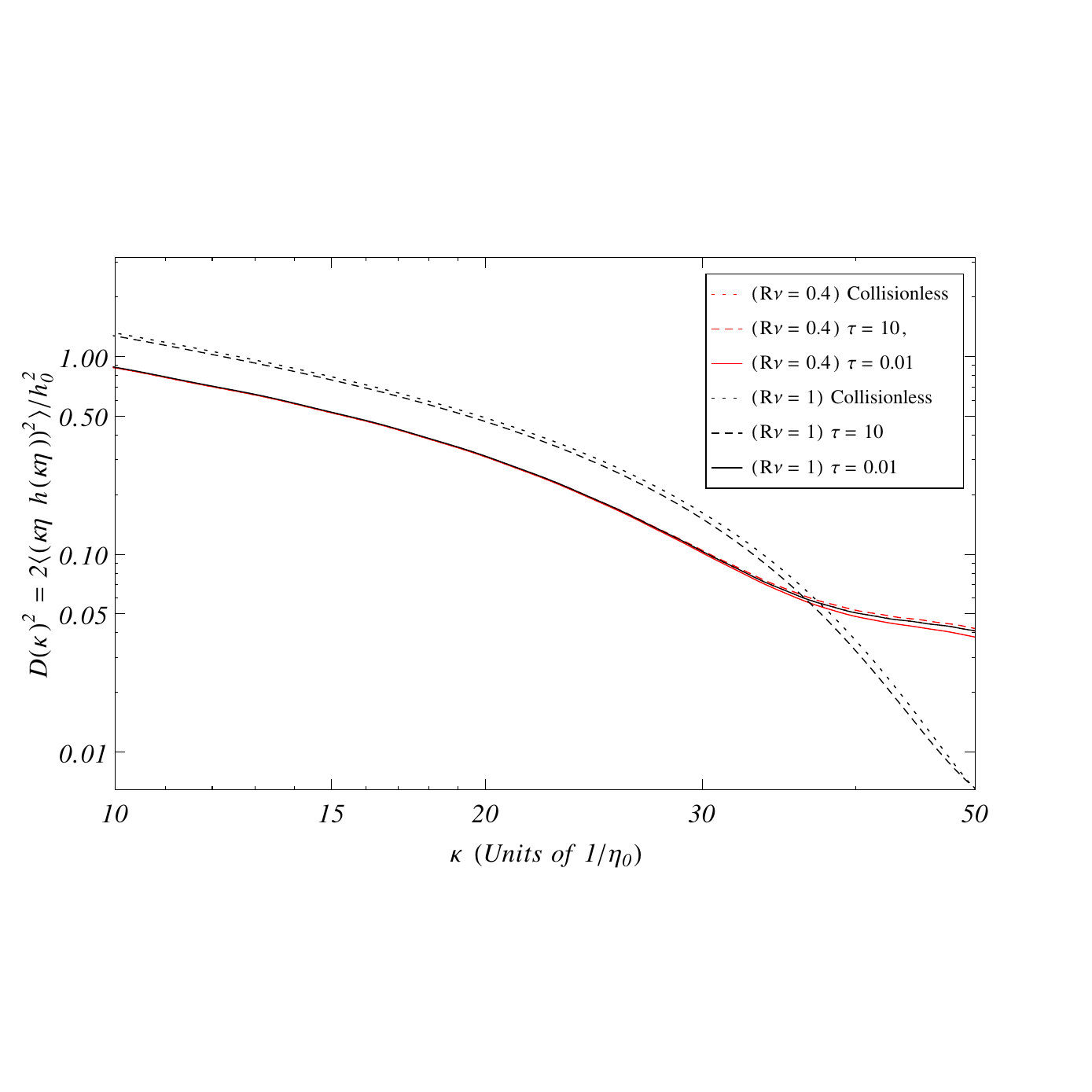}
\caption{Time-averaged values of $D(k)^2 = 2 \langle(k\eta)^2|h_{ij}(k\eta)|^2\rangle$ as function of $k\,[Mpc^{-1}]$ for the RMD scenario in case of including the collision term.
Results are for $\tau = 0.01$ and $10$ and for the collisionless according to the legend, with $\tau$ in units of $\eta_{0}$ and $k$ in units of $1/\eta_{0}$ with red and black lines in correspondence with those from Fig.~\ref{gw006a2}.}
\label{gw006b2}
\end{figure}

\begin{figure}
\centering
\includegraphics[scale=1]{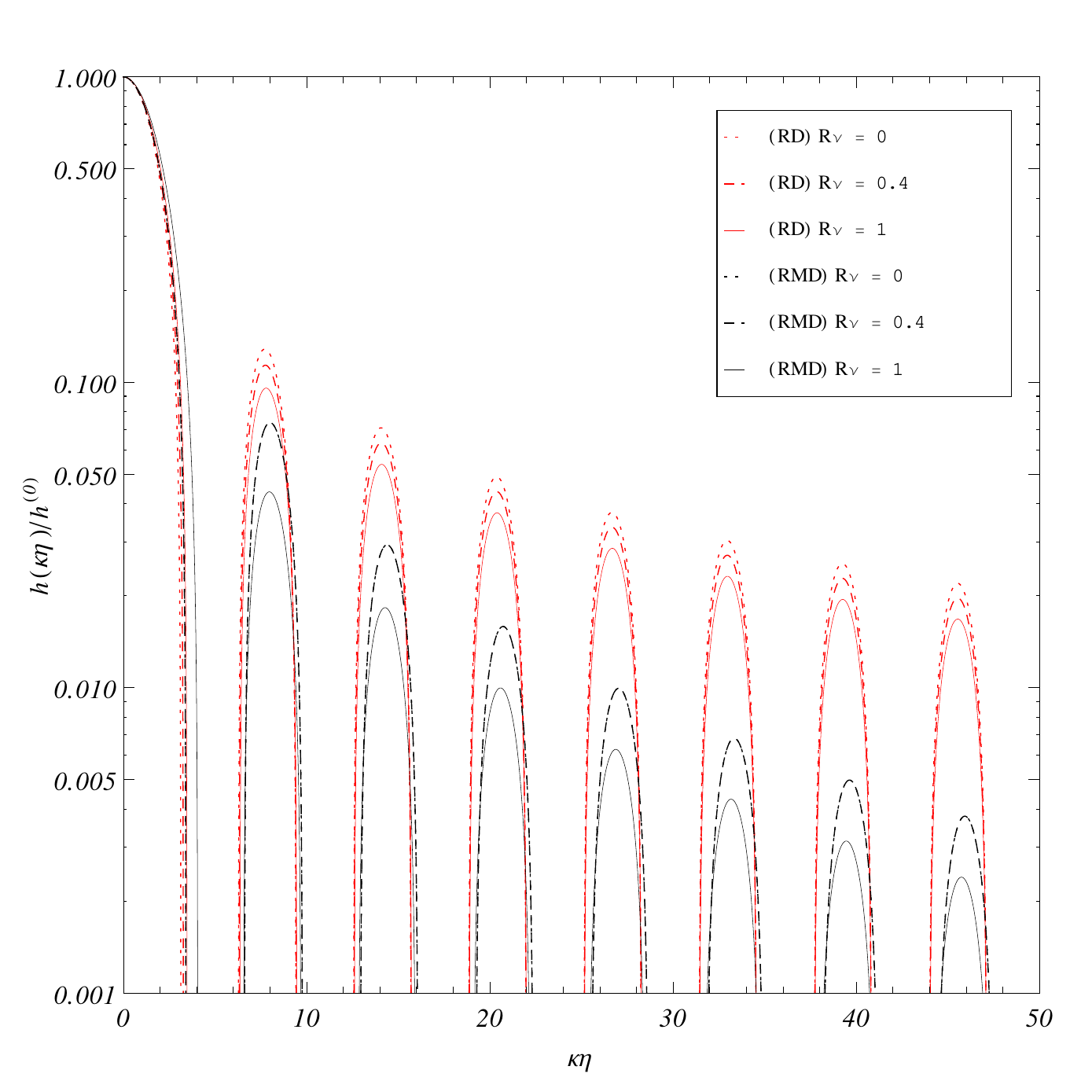}
\caption{Evolution of the normalized wave amplitude $h_{ij}/h_{ij}^{(0)}$ as function of $k \eta$ for RD (red lines) and RMD (black lines) background cosmic inventories, in correspondence with Fig.~\ref{gw006a1}.
Again, the RD curves are scale independent and the RMD curves are correctly interpreted by observing that $k \eta_{eq} = 100$, which correspond to scales that have entered the Hubble horizon before the time of matter-radiation equality.}
\label{gw006a1C}
\end{figure}

\begin{figure}
\centering
\includegraphics[scale=1]{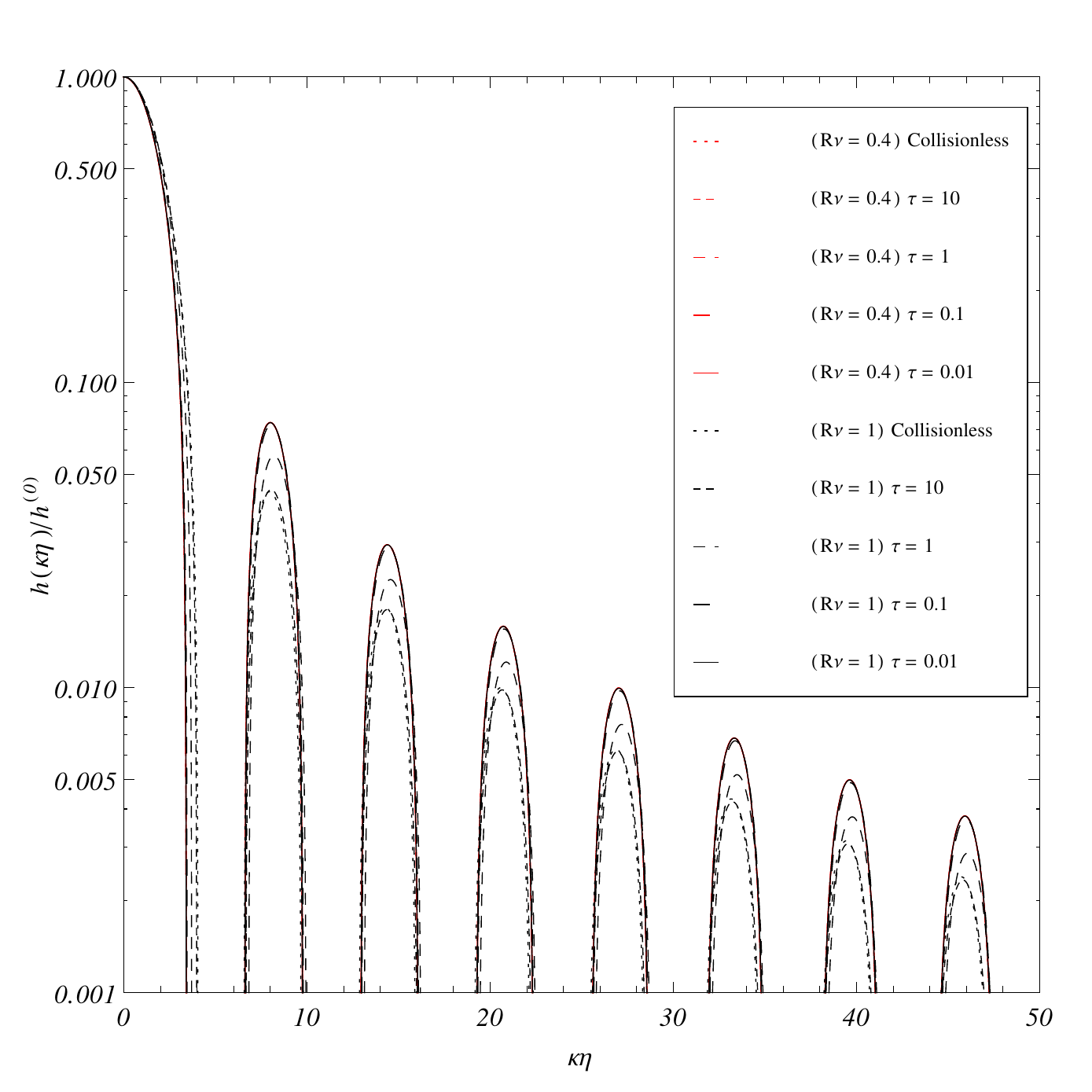}
\caption{Evolution of the normalized wave amplitude $h_{ij}/h_{ij}^{(0)}$ as function of $k \eta$ for the RMD scenario in case of including the collision term, in correspondence with Fig.~\ref{gw006a2}, with RMD curves being correctly interpreted by observing that $k \eta_{eq} = 100$.}
\label{gw006a2C}
\end{figure}

Figs.~\ref{gw006a1C} and \ref{gw006a2C} reveal the intrinsic dependence on the cross horizon driving parameter $k\eta_{eq}$.
They correspond to qualitative complementary results to Figs.~\ref{gw006a1} and \ref{gw006a2}, in case of considering $k\eta_{eq} = 100$ in place of $k\eta_{eq} = 1$.
For the RMD scenario, it corresponds to scales that have entered the Hubble horizon before the time of matter-radiation equality.

\begin{acknowledgments}
A. E. B. would like to thank for the financial support from the Brazilian Agencies FAPESP (grant 08/50671-0) and CNPq (grant 300233/2010-8).

\end{acknowledgments}

\vspace{10cm}
\end{document}